%% file: sample-sigconf.tex
\documentclass[sigconf]{acmart}

\usepackage{booktabs} 
\usepackage[english]{babel}
\usepackage{array}
\usepackage{tabularx}
\usepackage{booktabs} 
\usepackage{relsize}
\usepackage{rotating}
\usepackage[export]{adjustbox}
\usepackage{subcaption,siunitx,booktabs}
\usepackage{multirow}
\usepackage{caption} 

\setcopyright{none}

\renewcommand\footnotetextcopyrightpermission[1]{} 
\begin{document}
\title{Fighting Redundancy and Model Decay with Embeddings}

\author{Dan Shiebler}
\affiliation{
  \institution{Twitter Cortex}
}
\email{dshiebler@twitter.com}

\author{Luca Belli}
\affiliation{
  \institution{Twitter Cortex}
}
\email{lbelli@twitter.com}

\author{Jay Baxter}
\affiliation{
  \institution{Twitter Applied Research}
}
\email{jbaxter@twitter.com}

\author{Hanchen Xiong}
\affiliation{
  \institution{Twitter Applied Research}
}
\email{hxiong@twitter.com}

\author{Abhishek Tayal}
\affiliation{
  \institution{Twitter Cortex}
}
\email{atayal@twitter.com}
\begin{abstract}

Every day, hundreds of millions of new Tweets containing over 40 languages of ever-shifting vernacular flow through Twitter. Models that attempt to extract insight from this firehose of information must face the torrential covariate shift that is endemic to the Twitter platform. While regularly-retrained algorithms can maintain performance in the face of this shift, fixed model features that fail to represent new trends and tokens can quickly become stale, resulting in performance degradation. To mitigate this problem we employ learned features, or embedding models, that can efficiently represent the most relevant aspects of a data distribution. Sharing these embedding models across teams can also reduce redundancy and multiplicatively increase cross-team modeling productivity. In this paper, we detail the commoditized tools, algorithms and pipelines that we have developed and are developing at Twitter to regularly generate high quality, up-to-date embeddings and share them broadly across the company.

\end{abstract}

	\maketitle

\input{samplebody-conf}

\bibliographystyle{ACM-Reference-Format}
\bibliography{sample-bibliography} 
\end{document}

%% file: samplebody-conf.tex
\section*{Introduction}

Most Machine Learning algorithms operate on vectors. However, many entities' natural vector representation is very sparse or high-dimensional. For example, text is often represented as a series of sparse one-hot vectors, graphs are often written as highly-sparse adjacency matrices, and pixel representations of images and videos are high-dimensional and over-determined.

Unfortunately, many powerful models work best on low-dimensional dense entity representations. To address this mismatch, researchers have developed a variety of methods for generating embeddings or low-dimensional learned representations. These methods compress sparse or high-dimensional structure into small information-dense vectors. For example, matrix factorization and co-occurrence techniques like in \cite{word2vec} \cite{pca} \cite{pmf} extract low-dimensional latent vectors from large and sparse matrices. Furthermore, neural network-based techniques like \cite{imagenet} and \cite{autoencoder} extract nonlinear structure from redundant high -dimensional signals.

Recently, there has been a great deal of research on generating embeddings for arbitrary entities and employing them in production Machine Learning systems. For example, the authors of \cite{youtuberec} use entity embeddings to simplify the Youtube recommendation problem. In addition, the authors of \cite{starspace} describe an algorithm for generating generic entity co-embeddings. 


In this paper we introduce the embedding algorithms and tools that we are developing to simplify Machine Learning at Twitter. Our paper is organized as follows. First, we present some of the challenges that are unique to Twitter. Next, we describe our embeddings technical  stack, including our integrations with Airflow, a central Feature Registry, and a benchmarking system. We then outline a few Machine Learning tasks at Twitter that we solve through the application of the learned embeddings. Finally, we describe several of the embedding algorithms that we employ.

\section*{Unique Challenges of Twitter Data}
Twitter's platform presents a variety of unique challenges to any Machine Learning system. The amount of data to consume is massive (the so called \emph{firehose}): every day, Twitter's $300+$ million monthly active users author several hundred million Tweets. Further, because of Twitter's focus on real time information dissemination each Tweet needs to be processed and delivered to users within microseconds. 

Second, due to the myriad of text-based content, the presence of largely disjoint Twitter "interest groups", and Twitter's skewed follow graph (such that most nodes have only a few edges while some have tens of millions), most Machine Learning algorithms at Twitter naturally operate on sparse data. This can make training models particularly difficult \cite{sparseclassification} \cite{sparsetextcorpora} and force teams to rely on algorithms like binning or feature hashing \cite{featurehashing}. 

Third, the distribution of Twitter data is in constant flux. Topics that are 
trending now might essentially disappear in weeks, days, or even hours. For example, a specific event like an election or game may dominate conversations for a short period of time, and then die out. The detrimental effect of this kind of covariate shift on performance in described in \cite{creditcard} \cite{datasetshift}.

\begin{figure}[h]
\caption{Examples of different data shift in the Twitter ecosystem}
\includegraphics[width=0.5\textwidth]{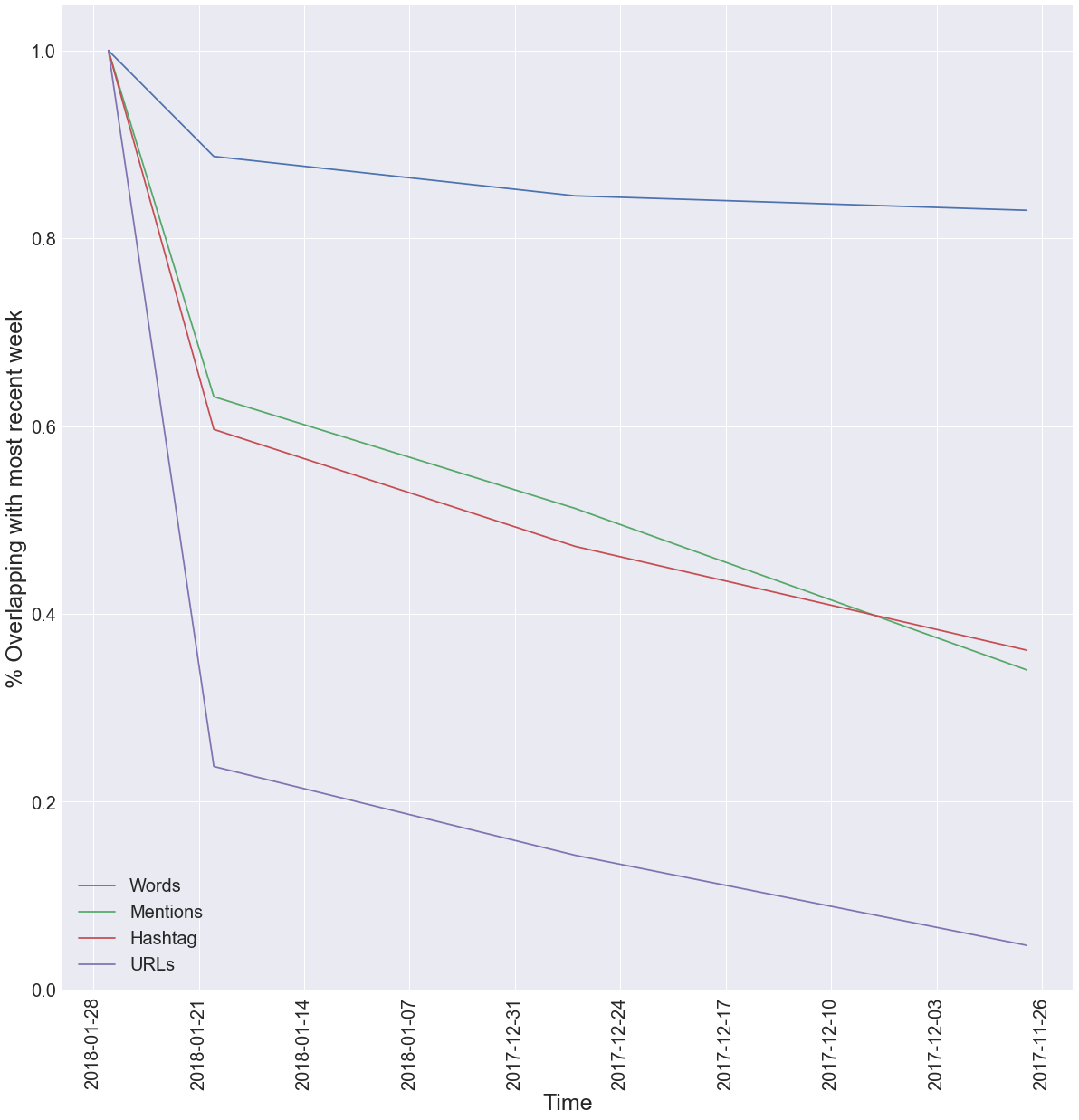}
\label{fig:DataShift}
\end{figure}

We demonstrate this effect by computing the percent overlap between this week's most popular 5K $@$ mentions, linked websites, hashtags and words, with those that were most popular last week, a month ago and a year ago (Figure \ref{fig:DataShift}). We observe that especially for mentions, linked websites, and hashtags, the frequency of overlap drops off quickly. A model trained a month ago will have been trained with less than 50\% of today's most popular hashtags.

\begin{figure}[h]
\caption{Word vector embeddings trained on Tweets written during different time periods show significantly different relationships between words.}
\includegraphics[width=0.5\textwidth]{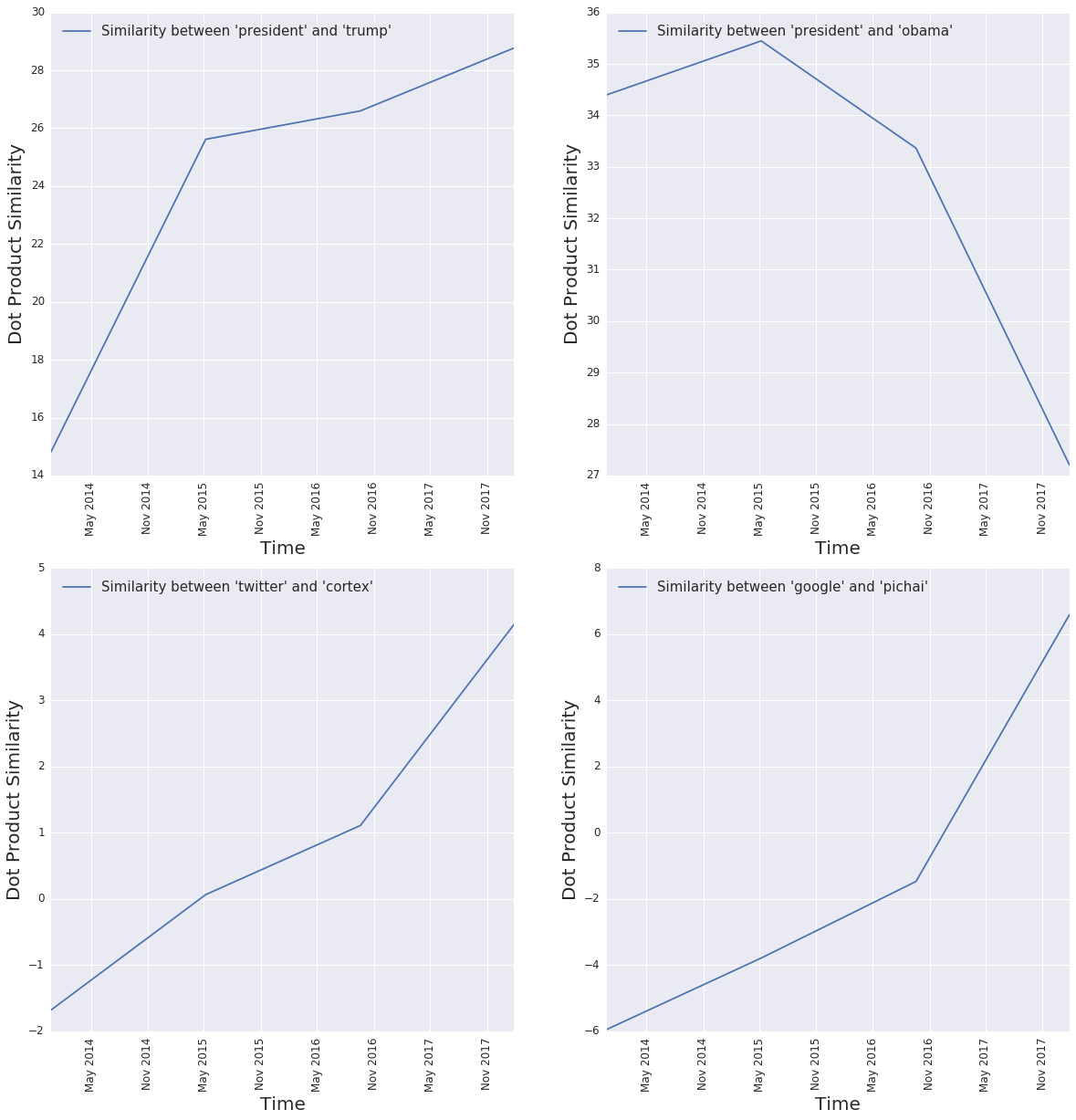}
\label{fig:Word2VecShift}
\end{figure}

In addition to this coverage dropoff, the meanings of words also change over time. This is not only a problem at Twitter: lingual shift is endemic to any text-based online platform. To see how this is problematic, consider a model that accepts text as inputs, where the text is represented as a series of word embeddings \cite{word2vec} \cite{w2vdownstream}. As words shift in meaning, word vector models become stale and degrade the performance of all models that depend on them. To illustrate this effect, we trained a series of skipgram word embedding models over the last several years of Tweet data and observed that certain word relationships changed dramatically over that time period. We show some clear examples in Figure ~\ref{fig:Word2VecShift}.

Luckily, we can employ entity embeddings to solve all three of these problems. First, since entity embeddings are low dimensional, they reduce the computational load on Machine Learning models across teams. Second, since entity embeddings are dense representations of sparse structure, they naturally work well with most Machine Learning approaches. Third, since regularly retrained embeddings naturally represent the most relevant aspects of data distributions, embeddings can reduce the impact of covariate shift on Twitter.

\section*{Our Embeddings Technical Stack}

Unlike hand-crafted features, which are generated by rule-based algorithms, learned embeddings are themselves outputs of models. These models must be trained from data, regularly retrained and benchmarked alongside the models that they are used with and served at scale. We are developing a series of tools that make it simpler for teams to customize, develop, access and share embeddings.

\subsubsection*{Workflows} 
~\\
At Twitter, we have integrated Airflow \cite{airflow} into our technical stack in order to manage complex Machine Learning workflows. Airflow allows us to replace the glued-together ad-hoc scripts that form most Machine Learning workflows into reusable programmatic components. Importantly, the various components can be very different in nature. For example, many embedding pipelines include both \emph{Scalding} (Twitter's Scala Map-Reduce library) for data collection and \emph{DeepBird} (Twitter's Python and Lua Deep Learning library) for training. By defining a pipeline and scheduling it to regularly execute, we are able to maintain fresh embedding models and win the battle against data distribution shift. In addition, establishing a "workflow" abstraction for all of the components of an embedding pipeline makes it easier for engineers to reason about the pipeline as a whole, rather than as a series of complex components. Furthermore, since each pipeline is built from modular components it's simple to adapt embedding generation pipelines into larger systems. In order to support end-to-end hyperparameter optimization, we have also integrated Airflow with Whetlab \cite{whetlab1}.
\\
To give an example, our Word2Vec pipeline consists of the following steps:
\begin{enumerate}
\item Execute a series of Scalding jobs to collect recent Tweets, concatenate them into conversations, identify commonly used words and phrases, and form skipgram pairs. 
\item Execute our \emph{DeepBird} Co-Occurrence Pipeline to generate word vector embeddings and publish them to the Feature Registry (See "Co-Embeddings"). 
\item Run benchmarking tasks on the trained embeddings and post the results. 
\end{enumerate}
This pipeline is fully automated and scheduled to run at regular intervals.

\subsubsection*{Feature Registry} 
~\\
We are in the process of rolling out Feature Registry, a central feature management store that allows teams to easily manage and share extracted features, including entity embeddings. It provides a unified access layer for any kind of raw, derived or learned feature (Figure 3), thereby abstracting the complexity of feature generation and streamlining the model construction and deployment process. Features for each Twitter entity, such as users and Tweets, are stored in the Feature Registry and made available for easy access by any model that operates on that entity.

\begin{figure}[h]
\caption{ Feature Registry abstracts feature complexity so even features like
embeddings can be easily used. }
\includegraphics[height=2.5in,  width=2.5in, keepaspectratio]{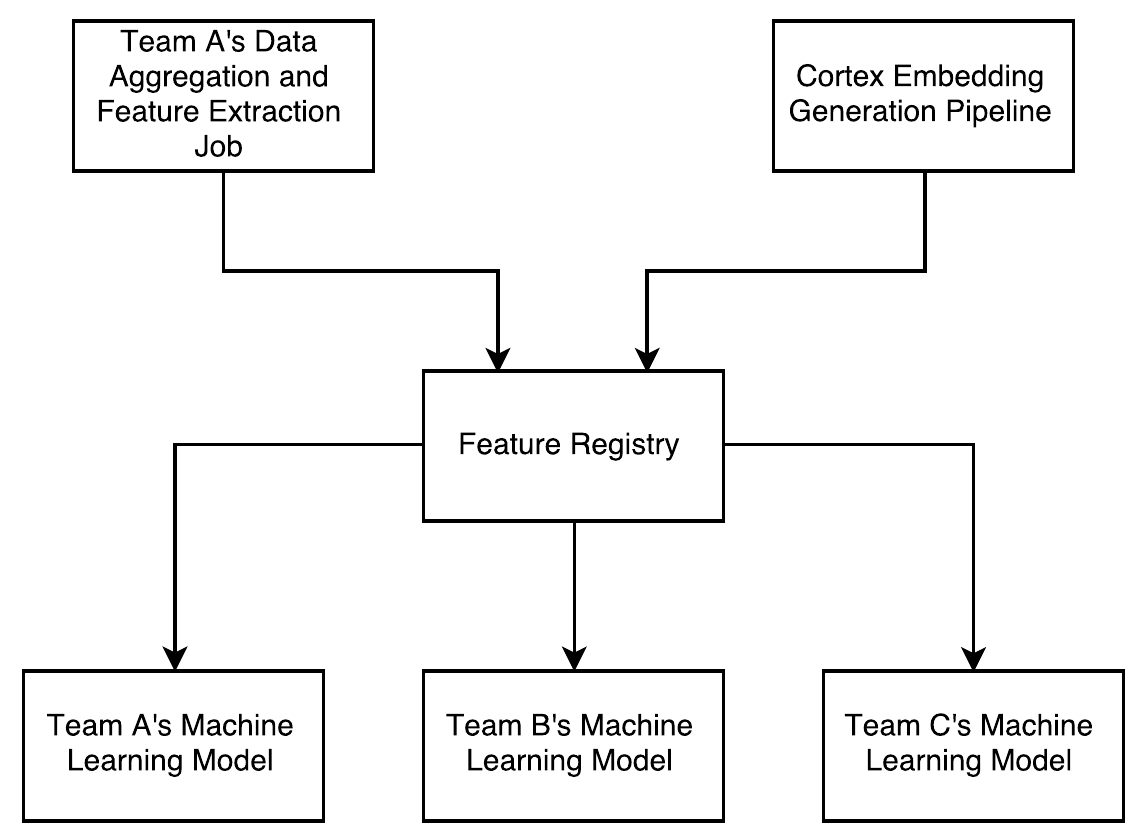}
\label{fig:FeatureRegistry}
\end{figure}


Feature Registry is also integrated tightly with Airflow, which allows us to automatically publish new versions of embeddings and their respective benchmarks to the Feature Registry. This kind of centralized feature sharing tool dramatically reduces the effort required for teams to develop Machine Learning pipelines.

\subsubsection*{Embedding Benchmarking System}
~\\
Unlike with a classification or regression model, it's notoriously difficult to measure the quality of an embedding. One of the reasons for this is that different teams use embeddings differently. For example, while some teams use user embeddings as model inputs, others use them in nearest neighbor systems. To mitigate this problem we have developed a variety of standard benchmarking tasks for each type of embedding. Every time an embedding is retrained it is automatically re-evaluated on these benchmarks, and the results are published in the Feature Registry along with the embeddings. Some examples of the task we include in our benchmarking are:
\begin{itemize}
\item \textbf {User Topic Prediction.} During onboarding, Twitter users may indicate which topics interest them. The ROC-AUC of a logistic regression trained on user embeddings to predict those topics is a measure of that embedding's ability to represent user interests.
\item \textbf {Metadata Prediction.} Certain users provide their demographic information (such as gender, age, etc). The ROC-AUC of a logistic regression trained on user embeddings to predict this metadata is a measure of how well that embedding might perform on a downstream Machine Learning task.
\item \textbf {User Follow Jaccard.} We can estimate the similarity of two users' tastes by the Jaccard index of the sets of accounts that the users follow. Over a set of user pairs, the rank order correlation between the users' embedding similarity (as determined with a similarity metric like Dot Similarity $uv$, Cosine Similarity $\frac{uv}{\|u\|\|v\|}$ or Euclidian Similarity $1 - \|u - v\|$) and their follow sets' Jaccard index is a measure of how well the embedding groups users.
\end{itemize}
See Table ~\ref{table:benchmarkingtasks} for some embeddings' results on these tasks. 

\section*{Example Tasks}
We now introduce two tasks that we will refer back to several times throughout the rest of the paper as concrete examples of tasks on which embeddings can improve performance. We also now introduce the strategies we use to incorporate embeddings into our solutions to these tasks. 

\subsection*{Tweet Email Recommendations}
One of the most important objectives at Twitter is motivating existing users to enter the platform and interact with new Tweets. One of the best ways to do this is to identify which Tweets a user might find interesting and email those Tweets to that user. An "Email Recommendation" is considered successful if the user chooses to click the Tweet in the email. Therefore the Tweet Email Recommendation problem is: given a user and a Tweet, determine whether the user would click on this Tweet in an email.

\subsection*{New User Follow Recommendations}
The very first experience users have after creating a Twitter account is the New User Experience (NUX), where users are asked to upload an address book, select their interests, and then receive recommendations for accounts to follow. Since the accounts followed from this step compose a user's entire original timeline, there is a large potential for these recommendations to make or break the initial Twitter experience. In addition, since this recommendation must occur immediately after a user signs up, many of the most important signals (such as the Tweets a user likes or the other accounts they follow) are absent. We will focus on one important submodule of the overall recommendation algorithm: Follow Prediction. Given the information about a user that is present at signup time and a potential account to follow, predict whether the user will choose to follow this account (and not unfollow shortly after signup).

\subsection*{Solving Tweet Email Recommendations and New User Follow Recommendations}

\begin{table*}[t]
\centering
\begin{tabular}{@{\extracolsep{4pt}}llccccccc}
& 100 Element TFW ALS & 300 Element TFW Co-Occurrence & 1000 Element SVD & 50 Element Autoencoded SVD \\ 
\midrule
User topic Prediction 
& ROC-AUC: 0.626 
& ROC-AUC: 0.656
& ROC-AUC: 0.813
& ROC-AUC: 0.786
\\
Metadata Prediction 
& ROC-AUC: 0.798
& ROC-AUC: 0.818
& ROC-AUC: 0.827
& ROC-AUC: 0.772
\\
User Follow Jaccard
& Spearman $\rho$: 0.314
& Spearman $\rho$: 0.298
& Spearman $\rho$: 0.522
& Spearman $\rho$: 0.356
\\
\bottomrule
\end{tabular}
\caption{The performance of various embedding methods on our Embedding Benchmarking System tasks.} 
\label{table:benchmarkingtasks}
\end{table*}

\begin{table*}[t]
\centering
\begin{tabular}{@{\extracolsep{4pt}}llccccccc}
& Baseline & Baseline + 1000 Element SVD Embedding  & Baseline + 50 Element Autoencoded SVD Embedding  \\ 
\midrule
User bucket 1 & 0.9252 & \textbf{0.9258} & \textbf{0.9258} \\
User bucket 2 & 0.8741 & \textbf{0.8751} & \textbf{0.8751} \\
User bucket 3 & 0.8458 & \textbf{0.8465} & 0.8464 \\
User bucket 4 & 0.9408 & \textbf{0.9413} & 0.9412 \\
User bucket 5 & 0.8696 & 0.8703 & \textbf{0.8704} \\
User bucket 6 & 0.8941 & \textbf{0.8952} & 0.8950 \\
User bucket 7 & 0.7796 & \textbf{0.7845} & 0.7830 \\
User bucket 8 & 0.9292 & 0.9304 & \textbf{0.9319} \\
User bucket 9 & 0.8640 & \textbf{0.8649} & 0.8648 \\
Average & 0.9341 & \textbf{0.9345} & 0.9344 \\
\bottomrule
\end{tabular}
\caption{ROC-AUC on the Tweet Email Recommendations task using a baseline model versus using the wide-and-deep model to incorporate user embeddings. We observe that adding user embeddings creates a consistent performance improvement. User are divided by bucket based on their level of interaction with the platform.} 
\label{table:emailrecommendation}
\end{table*}

\begin{figure}[h]
\caption{The wide and deep model accepts both embeddings and conventional sparse features.}
\includegraphics[height=2.5in,  width=2.5in, keepaspectratio]{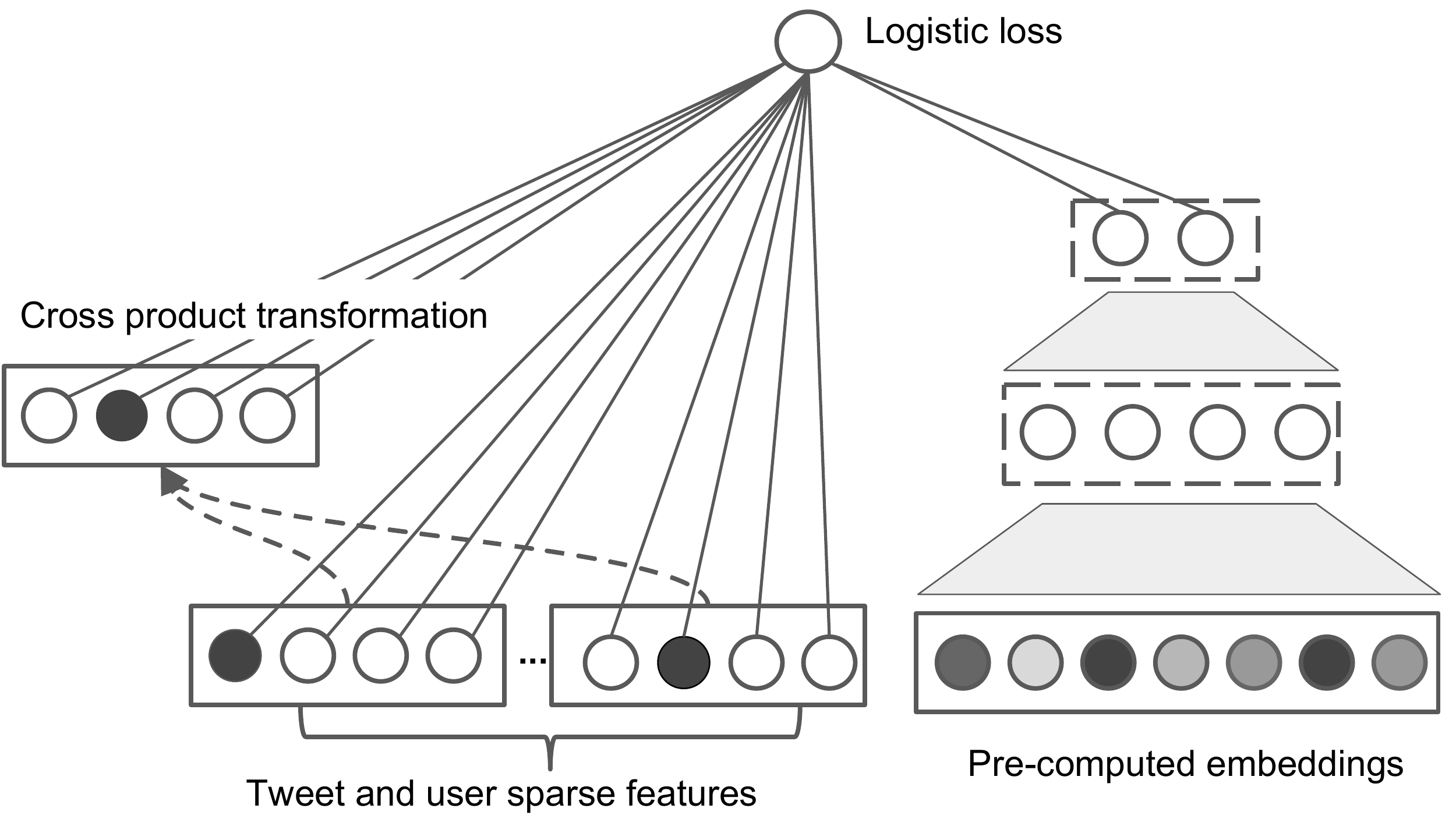}
\label{fig:wide-and-deep}
\end{figure}

A simple solution to both the Tweet Email Recommendation and New User Follow Recommendation problems is to use a model like a multi layer perceptron (MLP) that accepts user and Tweet/Account-to-Follow features. These features might include include descriptive data like Tweet length, creator information, existing engagement statistics, user metadata, \emph{etc}. In order to eke the maximal performance out of these features, we applied techniques like Feature Hashing, MDL and the sparse cross-product transformation \cite{featurehashing} \cite{MDL}. In order to evaluate the degree to which embeddings can improve the performance of these models, teams at Twitter have developed versions of the wide-and-deep (WAD) model \cite{wide-and-deep} that accepts both conventional sparse features and pre-computed embeddings. The model architecture is presented in Figure \ref{fig:wide-and-deep}. Note that since the conventional sparse features are already very heavily engineered for their respective problems, providing additional improvements on top of these features with a generic embedding model is not a simple task. 

\begin{table*}[t]
\centering
\begin{tabular}{@{\extracolsep{4pt}}llccccccc}
Features & RCE & ROC-AUC \\ 
\midrule
Baseline & 24.84 & 0.815 \\
Baseline + 100 Element ALS TFW Embedding & 27.56 & 0.835 \\
Baseline + 100 Element ALS TFW Embedding + 1000 Element Producer SVD & 27.88 & 0.836 \\
\bottomrule
\end{tabular}
\caption{Impact of the ALS TFW Embedding and SVD Producer embedding on NUX model performance. RCE is the relative cross entropy, which reports the percent improvement in cross entropy loss (on the validation set) between the model and a baseline model which always predicts the mean.}
\label{table:NUXResultsALS}
\end{table*}
\section*{Embedding Strategies}
In the next several sections we describe the techniques that we use to embed entities at Twitter.



\subsection* {Matrix Factorization}
One of the most popular approaches for representing users in terms of their interactions with items (\emph{e.g.} Tweets, other users) is the Low Rank Matrix Factorization paradigm. In this approach we form the sparse user-item affinity matrix and approximate it as the product of a low-rank user matrix and a low-rank item matrix. The reasons for Matrix Factorization's popularity are easy to understand: it's simple to interpret (each dimension within the low-rank embeddings can be interpreted as an underlying "latent factor") and there exist a variety of high-quality implementations that are suitable to large datasets \cite{sparkSVD} \cite{randomSVD} \cite{mapreduceALS}. We now introduce two examples of Matrix Factorization's successes at Twitter.
\begin{table*}[t]
\centering
\begin{tabular}{@{\extracolsep{4pt}}llccccccc}
& LOS-Accounts & Known-For & Interested-In  \\ 
\midrule
Follow Graph DeepWalk \cite{deepwalk} & 0.684 & 0.675 & 0.859 \\
Follow Graph SVD & 0.757 & 0.721 & 0.880 \\
\bottomrule
\end{tabular}
\caption{Evaluation of the SVD Producer embeddings versus a baseline follow graph embedding on three benchmark tasks.}
\label{table:SVDResultsTable}
\end{table*}

\subsubsection*{Consumer-Producer Engagement Matrix Factorization}
~ \\
On Twitter, users are classified into one of two roles based on their behavior patterns: \emph{Producers} and \emph{Consumers}. Producers are users who have relatively large numbers of high quality followers, \emph{e.g.} celebrities, activists, or politicians, while Consumers are all other users. By gathering and organizing engagement data (e.g. Likes, Retweets) between Consumers and Producers, we can form the sparse affinity matrix $X$ where each entry $x_{ij}$ represents the engagement strength of a Producer $j$ to a Consumer $i$. To get rid of noise, reduce the matrix sparsity and improve efficiency, we prune the graph with following steps: 
\begin{enumerate}
\item Get the set of all follows from normal active users and find the 1 million users that have the most follows in this set. This is the set of producers. 
\item Form $X_{raw}$ from the set of engagements between normal active users and these producers.
\item Compute the row and column sums of $X_{raw}$ and use these sums to normalize each element $x_{i,j}^{raw}$:
\begin{equation}
	x_{i,j} = \frac{x_{i,j}^{raw}} {\sqrt{\sum_{c \in cols}{x_{i,c}^{raw}}}* \sqrt{ \sum_{r \in rows}{x_{r,j}^{raw}}}}
\end{equation}
Although normalizing the rows of a matrix before performing SVD is a common practice \cite{Herlocker1999} we found that applying normalization to both the columns and rows helped reduce the impact of Twitter's heavily lopsided follow graph and improve performance. 

 
\end{enumerate}

After the above construction and pruning steps, we can perform singular value decomposition (SVD) on the normalized matrix $X$:
\begin{equation}
	X = U\Sigma V^{T}
\end{equation}
Where $\Sigma$ is the diagonal singular value matrix and the columns of $U$ and $V$ are the left and right singular vectors respectively. To obtain a low-rank factorization of $X$, we take the top-k singular values and associated left/right singular vectors. Since the magnitude of singular value $i$ reflects the significance of both the $i^{th}$ left and the $i^{th}$ right singular vectors, it's natural to absorb the square root of singular values into $U$ and $V$ to form the Consumers' and Producers' embedding matrices $U^*$ and $V^*$: 
\begin{equation}
   X = \left(U\sqrt{\Sigma}\right) \left(\sqrt{\Sigma}^T V^T\right) = U^* V^{*^T}
\end{equation}

In Table \ref{table:emailrecommendation} we demonstrate that incorporating the user embeddings that we generate with this technique consistently improves the performance of the Email Recommendation model across all levels of user activity. In Table \ref{table:NUXResultsALS}, we demonstrate that incorporating Producer embeddings into the NUX model can improve performance over using just a user embedding. 

In order to quantify the performance of these Producer embeddings in a more direct manner, we also compute user embeddings with the DeepWalk algorithm \cite{deepwalk} and evaluate the results on three tasks with human labeled data:
\begin{enumerate}
\item \textbf{LOS-Accounts}: In this task, we train a logistic regression model on the top 10,000 producer embeddings to classify them into one of 59 human-determined interest categories. We measure the model's performance as its classification accuracy. 
\item \textbf{Known-For} and \textbf{Interested-In}: In these tasks, for each of the top 100,000 producers we train a multi-output linear regression model on that producer's embedding to predict the degree to which that producer is respectively "known for" and "interested in" each of 6011 tags (\emph{e.g}. "news", "hollywood", "gastronomy", "women in science", etc.). We measure the model's performance as the average value of the NDCG between the model induced and human labeled rankings.
\end{enumerate}

On all three tasks we find that our SVD user embeddings outperform the DeepWalk embeddings (Table \ref{table:SVDResultsTable}).

\subsubsection*{TFW Interaction Matrix Factorization}
~ \\

Before many browsers are used to sign up for Twitter, they have already interacted with Twitter on websites with embedded Twitter content, known as Twitter for Websites (TFW) domains. Similar to Consumer-Producer engagement embeddings, we can learn TFW embeddings by factorizing the browser-TFW interaction matrix. Since this data is available during the NUX, TFW domain embeddings can be particularly useful for the NUX Follow Recommendation task, where few other strong signals exist.

An interesting facet of TFW data is that it's extremely skewed: over 80\% of browser instances interact with the top 5 or so domains, but the distribution drops off quickly so that less than half a percent of browser instances interact with the $100^{th}$ most popular domain. We find that these most popular TFW domains (\emph{e.g.} www.google.com) are mostly uninformative, so we remove them. This leads to extreme sparsity, so we use the Alternating Least Squares (ALS) approach from \cite{implicit} to downweight the impact of $0$s on the matrix factorization objective. Our algorithm consists of the following steps:
\begin{enumerate}
\item Select the 1 million most popular TFW domains and gather the interactions between browser instances and these domains to form the matrix $A$ of browser-TFW interactions. Normalize this matrix similarly to how we normalize the Consumer-Producer matrix. 
\item Apply the Alternating Least Squares algorithm to factorize this matrix into the browser and domain matrices $U$ and $V$ such that $A \approx UV$.
\end{enumerate}

As we can see in Table \ref{table:NUXResultsALS}, the ALS TFW domain embeddings provide a significant performance improvement for the NUX task.

\subsection*{Co-Embeddings}
\begin{figure}
\includegraphics[height=3in, width=3in, keepaspectratio]{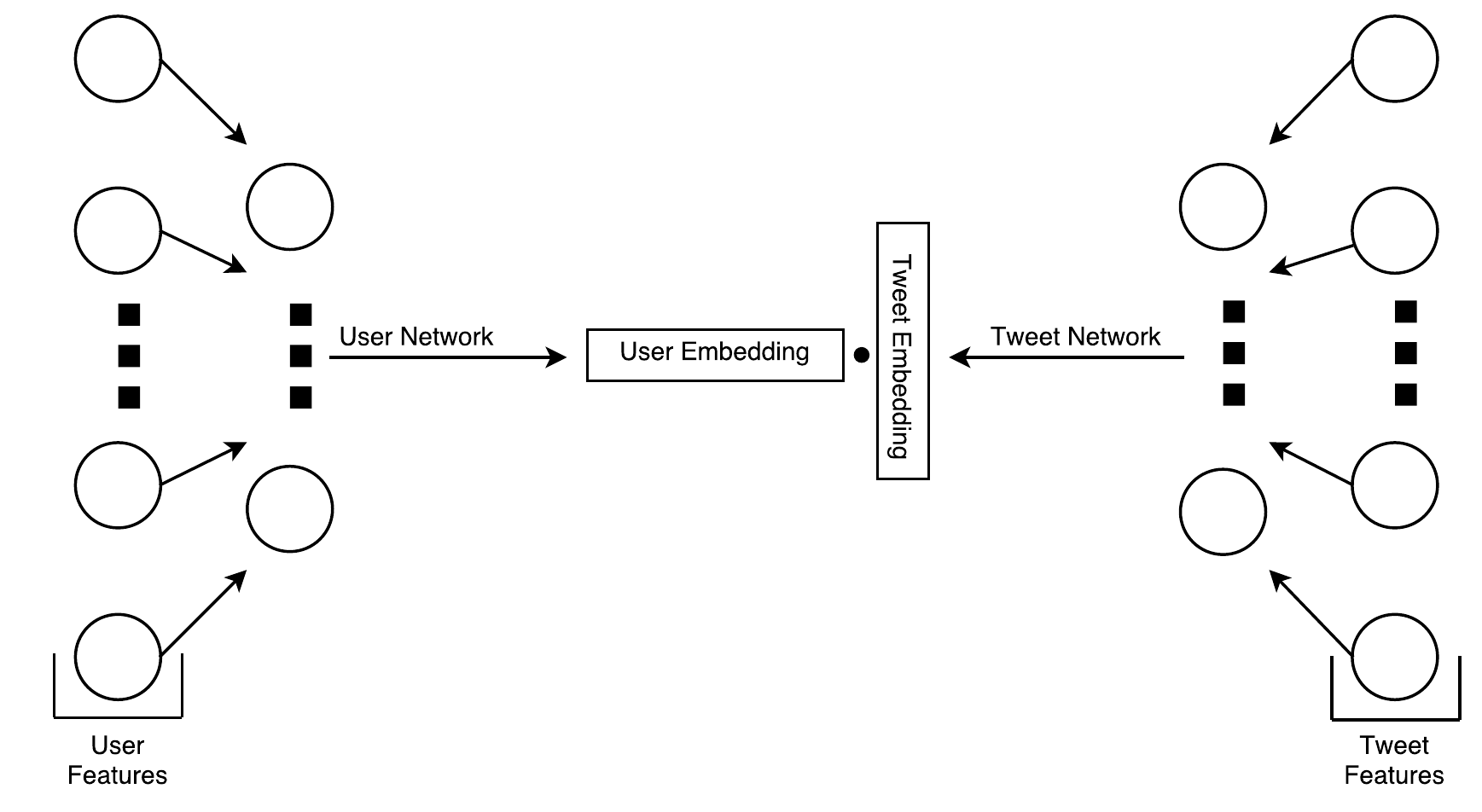}
\caption{A network to directly co-embed users and Tweets.
}
\label{fig:Coembedding}
\end{figure}

Most product teams at Twitter deal with the basic underlying problem of selecting a small subset of entities (Tweets, users, events, etc) from a possible large set and presenting a ranked list of these entities to clients such that they deliver the desired customer experience. A few examples are:
\begin{enumerate}
\item The home timeline presents a ranked list of Tweets from a host of possible Tweets by first narrowing them down to Tweets engaged or authored by other users in your network or from out-of-network authors based on your interests and past engagements. 
\item Ads selects a small ranked list of line items that the customer is most likely to engage with from a possible large set, eligible to be displayed to the customer based on advertiser's chosen targeting criteria. 
\end{enumerate}

Although it would require a great deal of work for each team to develop their candidate recommendation systems in isolation, if we have embeddings for users and entities such that a similarity metric like dot product $uv$, cosine similarity $\frac{uv}{\|u\|\|v\|}$ or euclidian similarity $1 - \|u - v\|$ is indicative of user-item affinity, then this reduces to an approximate nearest neighbor problem. As such, an important effort we have made at Twitter is developing reliable user-item co-embeddings, and we are currently in the process of developing a large scale nearest neighbor system to use these co-embeddings for candidate recommendation. In the meantime, we have seen strong results by using co-embeddings directly in Machine Learning models. 

Note that the Consumer-Producer matrix factorization algorithm that we discussed above is an example of a co-embedding algorithm, where the Consumer and Producer embeddings are engineered such that the dot product of some Consumer's embedding and some Producer's embedding will be as close as possible to the number of engagements between that Consumer and that Producer. This framework works well when we are interested in a fully collaborative co-embedding algorithm without user or item metadata \footnote{Certain matrix factorization approaches can incorporate user/item metadata, but we omit them for this discussion} and we have a concrete measure of "affinity" between a user and an item that we want dot products to represent.

In order to generate user-item co-embeddings from arbitrary features, we are developing a generic co-embedding network system (similar to \cite{microsoftdeepcrossdomain}). Our system consists of two embedding networks, a user network and an item network, which accept feature representations of users and items respectively. The networks produce embeddings of the same length such that the dot product (or other similarity metric) between a user's and item's embeddings is indicative of the affinity between that user and item (See Figure ~\ref{fig:Coembedding} for an example).
This approach works well when we are interested in generating co-embeddings from complex nonlinear combinations of user and item features and we need generate realtime embeddings for new users and items based on these features.

\begin{table*}[t]
\centering
\begin{tabular}{@{\extracolsep{4pt}}llccccccc}
& Matrix Factorization & Co-Embedding Network & Co-Occurrence (Direct) & Co-Occurrence (Bag of Features)  \\ 
\midrule
Kind of Feature & Collaborative & User/Item Metadata & Collaborative & Feature Collaborative\\
Nonlinear & No & Yes & No & No \\
Scalability & Medium & Very High & High & Very High\\
Data Objective & User-Item Affinity & User-Item Affinity & Item-Item Co-Occurrence & Feature-Item Co-Occurrence\\
Handles New Users & Yes (with Folding-In) & Yes & No & Yes\\
Handles New Items & No & Yes & No & Yes \\
\bottomrule
\end{tabular}
\caption{A comparison of co-embedding strategies.} 
\end{table*}
In both of the above methods, our goal is to co-embed entities such that the similarity between two entities' embeddings is as close as possible to some measure of affinity between the entities. However, it's often more convenient to frame the interaction between two entities in terms of "co-occurrences". For example, a co-occurrence between a user and a Producer could be an instance of a user liking that Producer's content. Then the objective is to generate embeddings such that the similarity between two entities' co-embeddings is indicative of the entities' co-occurrence likelihood. 

At Twitter Cortex, we have developed a generic "Co-Occurrence Embedding" system (similar to \cite{starspace}) that can generate co-embeddings for entity types $e_1$ and $e_2$ from a set of $(e_{1i}, e_{2j})$ co-occurrence pairs. The pipeline integrates with Airflow, so engineers can implement workflows to regularly generate new co-occurrence pairs and retrain the embeddings (such as the word vector embedding workflow described in "Workflows"). The system performs the following algorithm to pull the embeddings for entities that tend to co-occur closer together, and push embeddings for entities that rarely co-occur apart:
\begin{enumerate}
\item Construct the embedding matrices $E_1$ and $E_2$, where the $i^{th}$ row of $E_1$ ($E_{1_i}$) and the $j^{th}$ row of $E_2$ ($E_{2_j}$) correspond to the embeddings for $e_{1i}$ and $e_{2j}$ respectively. 
\item For each pair of entities $(e_{1i}, e_{2j})$, select a group of "negative samples" $S_{e_{2}}$ from the set of $e_2$ entities and perform an SGD step to maximize the function $log \sigma(E_{1_i}E_{2_j}^{T}) + \sum\limits^{k \in S_{e_{2}}} log \sigma(- E_{1_i} E_{2_{k}}^{T})$
\end{enumerate}

There are three ways that we can use this pipeline to generate entity embeddings:
\begin{enumerate}
\item We can co-embed two entity types based on a co-occurrence criteria between them, such as the Consumer-Producer example above.
\item We can co-embed entity types $e_1$ and $e_2$ by representing $e_1$ as a "a bag of features", defining a co-occurrence criteria between $e_1$'s features and $e_2$, and assigning $e_1$'s embeddings to be the weighted average of it's feature embeddings.
\item We can embed a single entity type according to a co-occurrence criteria. For example, in \cite{word2vec}, Mikolov et al used this strategy (known as skipgram in this context) to generate word embeddings based on the co-occurrence criteria that two words appear near each other in a document. 
\end{enumerate}

Use case (2) is particularly robust in situations where we need to model new items in realtime, since we can compute new embeddings with just a table lookup and a vector average. For example, we can use (2) to co-embed users and Tweets by representing Tweets as "bags of words," and defining the user-word co-occurrence criterion as "word appears in Tweet that user likes." This strategy allows us to quickly compute new Tweet embeddings that we can directly match with existing user embeddings. 

We can also use (2) to generate embeddings for new users based on the TFW domain interactions of corresponding browsers. If we define a (TFW domain, Producer) co-occurrence event between domain $r$ and Producer $p$ as an instance of a user who follows $p$ and whose corresponding browser visits $r$, we can use this to generate embeddings for new users that we can easily match with existing Producers. We find that we can use this strategy to improve the performance of the NUX recommendation model. Just adding the dot product between $300$ element TFW domain embedding and the Producer's embedding to the feature set improves the RCE and ROC-AUC of the NUX model by $0.14$ and $0.001$ respectively. We can also see that these "TFW Co-Occurrence" embeddings perform well on the embedding benchmarking tasks (Table ~\ref{table:benchmarkingtasks}).


\subsection*{Folding-In New Entities}
One of the primary downsides of some of the embedding techniques that we have discussed (such as Matrix Factorization and Direct Co-Occurrence) is that all of the embeddings are generated and saved at the same time. This makes it difficult to assign embeddings to new entities without retraining the model from scratch and can even decrease model performance.
For example, consider the problem of generating Consumer and Producer user embeddings with a matrix factorization approach. The model uses the recent interactions between Consumers and Producers to quantify Consumers' affinities for Producers. For relatively mature and active users this is a good approximation. However, new and inactive users have significantly fewer interactions with Producers, so their Producer interactions are noisier approximations of their Producer affinities. Therefore, we may see better overall performance by omitting these noisy users from the model training.

To address these problems, we use "Folding-In" strategies to assign static embeddings to new entities without affecting the original embedding model. To illustrate how this works, consider a matrix factorization model where we approximate the user-item interaction matrix $X$ with the product of the low rank user matrix $U$ and the low rank item matrix $V$. Then for some new user with interaction vector $x$, we want to assign to them the embedding $u$ such that $\|uV - x\|$ is minimized. If our matrix factorization model is a vanilla SVD, this is equivalent to the problem of projecting the vector $x$ onto the user embedding vector space \cite{foldingin}. If we write our SVD such that
\begin{align}
X = U^{*}\Sigma V^{*^{T}} = (U^{*}\Sigma^{1/2})(\Sigma^{1/2}V^{*^{T}}) = UV
\end{align}
then we can project $x$ onto the row space of $U$ with
\begin{align}
u = xV^{{-1}} = xV^{*}\Sigma^{\frac{1}{2}^{-1}}
\end{align}
If our matrix factorization model is a more general least squares model along the lines of \cite{implicit} or \cite{svdpp}, then we don't have the orthonormality guarantees that SVD provides, so it's not quite as easy to perform this projection. However, we can still apply a least squares solution method or use an approximation of $V^{-1}$.

Since we can express the "Folding-In" procedure as the product of a sparse user-item interaction vector and a dense "fold-in" matrix, folding a new user in is as simple as querying Feature Registry for the rows of the dense matrix corresponding to the items that user interacted with and computing their sum weighted by the strength of the user's interactions with those items. This is valuable for two reasons. First, we can perform this operation online without any modeling architecture, making it an attractive approach for assigning embeddings in low latency settings. Second, in an offline setting we can express these operations in a Map-Reduce framework (such as Twitter's \emph{Scalding} library), which allows us to easily fold hundreds of millions of users into our matrix factorization models. 

During the map phase, the algorithm converts each item vector to a set of (item, index, value) tuples and joins them to the set of (user, item, interaction) tuples to form a set of (user, index, value * interaction) tuples. During the reduce phase, the algorithm sums along the user and index dimensions to produce the user embeddings. This structure allows the algorithm to easily parallelize across multiple machines and quickly assign embeddings to hundreds of millions of users. 

\subsubsection*{Folding-In Experiment}
~\\
In this experiment we collected a set of Twitter users and popular Twitter Producers, formed the sparse Consumer-Producer engagement matrix and explored how removing users with fewer engagements affected the performance of an Alternating Least Squares model \cite{implicit}.

\begin{figure}[h]
\centering
\includegraphics[width=0.5\textwidth]{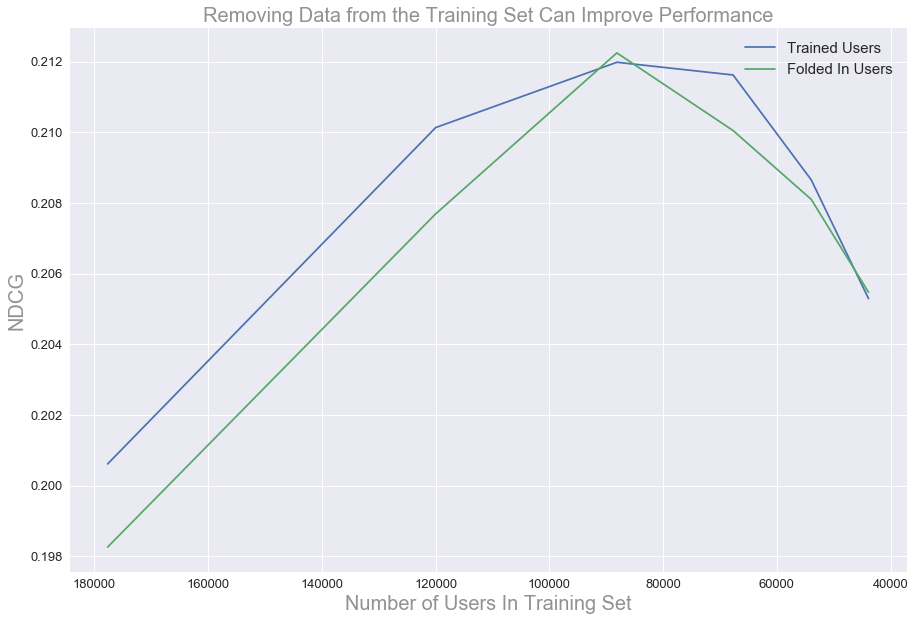}
\caption{An Alternating Least Squares model trained on Consumer-Producer interactions performs best when we remove the noisiest users from the training set and fold them into the model after.}
\label{fig:PercentData}
\end{figure}

To begin, we split our Consumer-Producer engagements into training and testing sets for each user. Then, we repeated the following algorithm for different values of $N$:
\begin{enumerate}
\item Select the top $N$ percent of users and use their training set engagements to train the Alternating Least Squares algorithm.
\item Assign user embeddings to the remaining users by multiplying their training set engagements with an approximation of $V^{-1}$.
\item Evaluate the model performance over the testing set engagements with the NDCG metric.
\end{enumerate}
We found that the model's performance initially increases as we remove noisier users from the training set, but it eventually peaks and begins to decrease after the training set becomes too small (Figure ~\ref{fig:PercentData}). Importantly, this effect holds over both the users who we fit the model on and the users who we fold into the model. That is, our predictions of noisy users' future engagements is more accurate when we leave them out of the model fitting stage than when we include them.

\subsubsection*{"Lookalike" Folding-In}
~\\
In certain situations we may want to use a co-embedding to predict the affinity between a user with no interaction data whatsoever and an entity. For example, we may want to determine the affinity between a new user and a Producer based on the Consumer-Producer SVD embedding. In these situations we can't use traditional Folding-In strategies, so we utilize a technique that we call "Lookalike Folding-In". 

\begin{figure}
\includegraphics[height=2in, keepaspectratio]{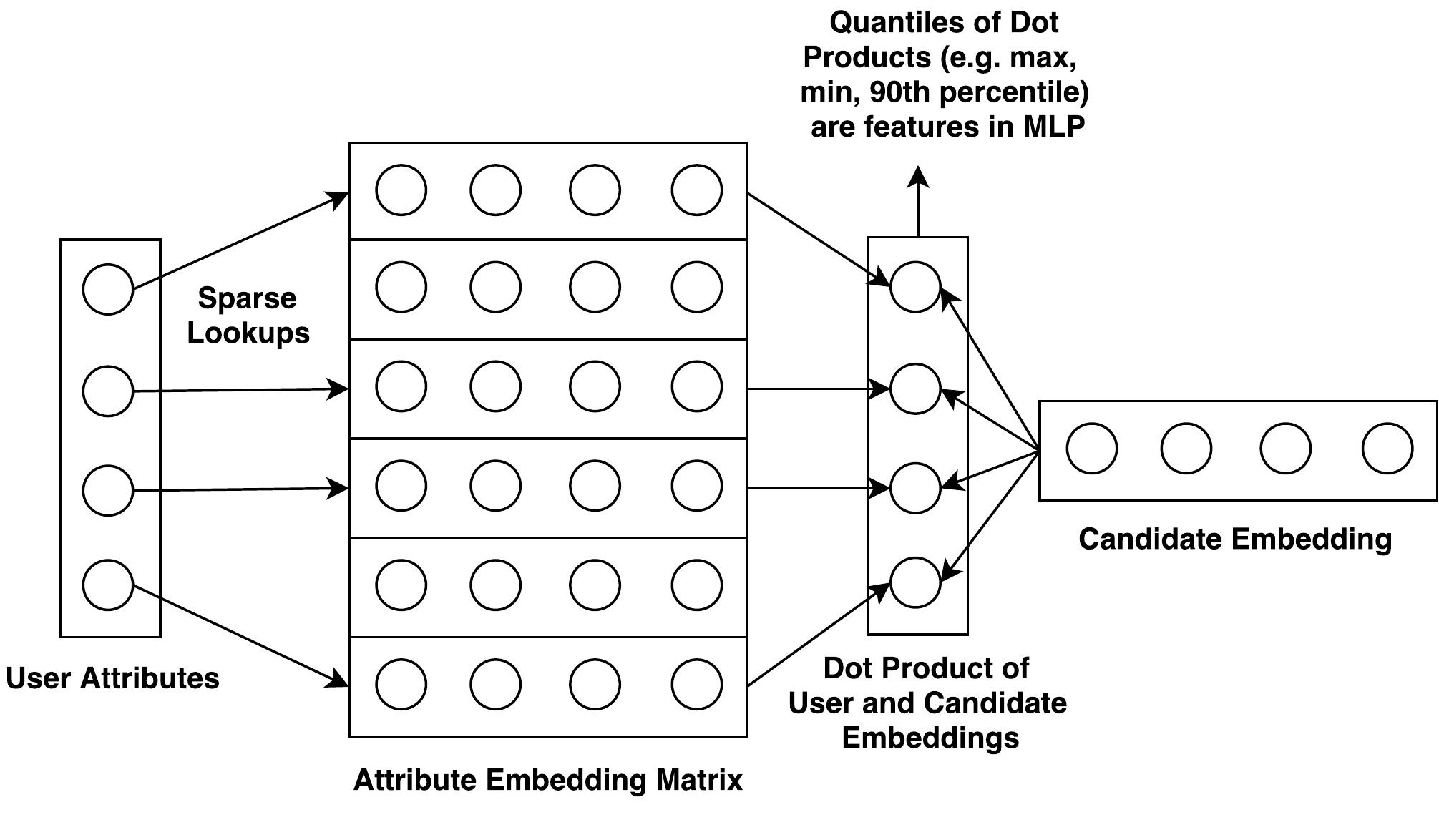}
\caption{Example of "Lookalike Folding-In" for new users.
}
\label{fig:SVDNewUserFeaturization}
\end{figure}

First, we look up user embeddings for existing users that are similar to the new user, \emph{e.g.} users in the new user's address book, users who selected the same interest categories when they signed up, and users who are in the same geographical area\footnote{For computational reasons, we average the user embeddings for each interest or geographical area.}. Next, we compute the similarity (dot product in this case) of each user embedding with the candidates' embeddings and we compute the quantiles (\emph{e.g.} max/median/min) of these similarity vectors to generate a set of user features that we can use in a model (Figure \ref{fig:SVDNewUserFeaturization}). 
We find that incorporating these features into the NUX model improves the RCE and ROC by $0.47$ and $0.004$ respectively. 

\subsection* {Democratized Embeddings}

\begin{figure}
\includegraphics[height=1.9in, keepaspectratio]{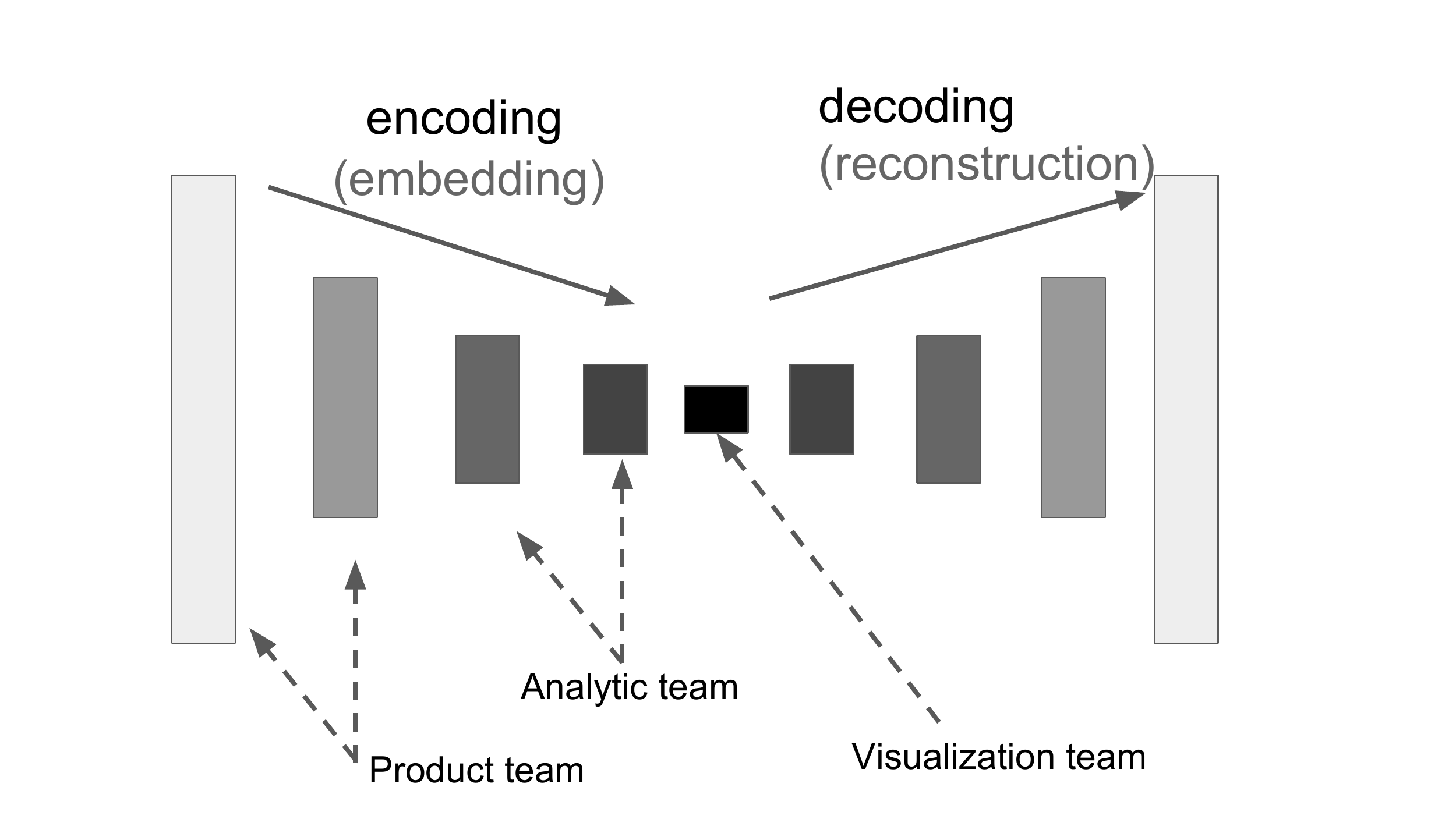}
\caption{We generate embeddings for different teams with a deep autoencoder.}
\label{fig:different_layers}
\end{figure}

Although multiple teams may be interested in the same embedding, it's not uncommon for different teams to have different efficiency and quality requirements and therefore be interested in embedding vectors of different lengths. For example, the analytics team at Twitter performs realtime interactive user segment analysis, which requires low latency modeling but is relatively robust to the lack of fine grained information about each user. Therefore, they would be most interested in very low dimensional user embedding vectors ($20$-$50$). In contrast, since the Email Recommendation team's models run offline latency is less of a concern, but their standards for performance are higher. Therefore, it would make sense for the Email Recommendation team to use higher dimensional user embedding vectors ($100$-$1000$).

Rather than resort to training separate embedding models for each team, we apply a variation on the \emph{Deep AutoEncoder} model \cite{deepAutoEncoder} to easily supply multi-length embeddings to teams throughout Twitter.  The \emph{Deep AutoEncoder} is composed of two symmetrical deep-belief networks. One network performs an encoding operation $f(\cdot)$ while the other one performs a decoding operation $g(\cdot)$ (Figure \ref{fig:different_layers}). The model accepts an original high-dimensional embedding as input, which it sequentially encodes into smaller and smaller embedding vectors, the final of which is decoded by the decoding network. We train the model to minimize the difference between the original input embedding and the decoded output: $\arg\min_{f,g} \|x - g(f(x))\|$. In our construction, we monotonically decrease the number of neurons from the input layer to the bottleneck layer, and we found that we achieve better performance by replacing the RBM's sigmoid activation function with the relu \cite{relu} activation function and training the model in an end-to-end fashion. 

Once this model is trained, each layer in the encoder can serve as a user representation of different dimensionality. As we demonstrate in Figure \ref{fig:different_layers}, each team can select the layer with the efficiency-quality tradeoff that best suits them. This decision is made particularly easy by the available of our Embedding Benchmarking System. We refer to this technique as "Embedding Democratization," and we demonstrate in Table ~\ref{table:emailrecommendation} and Table ~\ref{table:benchmarkingtasks} that using this technique to nonlinearly reduce the size of Consumer-Producer SVDs from $1000$ elements to $50$ elements does not dramatically impact the performance of the Email Recommendation pipeline (whereas simply taking the top $50$ elements of the SVD does dramatically reduce Email Recommendation performance).

\section*{Discussion}
In this paper we described the tools and algorithms that we have constructed at Twitter to facilitate generating and sharing embeddings. We also detail a variety of embedding methods, such as Matrix Factorization, Co-Occurrence, Folding-In and Democratization, and demonstrate the results of applying them to real world problems at Twitter.

There are few consistent guidelines and best practices for developing reusable Machine Learning systems, but we believe that tools and algorithms which facilitate modularity and collaboration - such as Workflows, Feature Registry, and the Embedding Benchmarking System - help mitigate Machine Learning's inherent pitfalls. By abstracting away the complexities of model building, these tools can allow Machine Learning to scale. 

\section*{Acknowledgements}
We would like to thank Matt Miller and Anthony Sciola for their work designing and building the large scale SVD embeddings, Min Lee for integrating embeddings into the email recommendation model, Hugo Larochelle for his evaluations of the SVD embeddings (work done while at Twitter), Nimalan Mahendran and Yi Zhuang for their helpful feedback, and Nico Koumchatzky for his support.

%% file: sample-sigconf.bbl

\begin{thebibliography}{28}


\ifx \showCODEN    \undefined \def \showCODEN     #1{\unskip}     \fi
\ifx \showDOI      \undefined \def \showDOI       #1{#1}\fi
\ifx \showISBNx    \undefined \def \showISBNx     #1{\unskip}     \fi
\ifx \showISBNxiii \undefined \def \showISBNxiii  #1{\unskip}     \fi
\ifx \showISSN     \undefined \def \showISSN      #1{\unskip}     \fi
\ifx \showLCCN     \undefined \def \showLCCN      #1{\unskip}     \fi
\ifx \shownote     \undefined \def \shownote      #1{#1}          \fi
\ifx \showarticletitle \undefined \def \showarticletitle #1{#1}   \fi
\ifx \showURL      \undefined \def \showURL       {\relax}        \fi
\providecommand\bibfield[2]{#2}
\providecommand\bibinfo[2]{#2}
\providecommand\natexlab[1]{#1}
\providecommand\showeprint[2][]{arXiv:#2}

\bibitem[\protect\citeauthoryear{??}{air}{2015}]%
        {airflow}
 \bibinfo{year}{2015}\natexlab{}.
\newblock \bibinfo{title}{Airflow}.
\newblock   (\bibinfo{year}{2015}).
\newblock
\showURL{%
\url{https://airflow.apache.org/}}


\bibitem[\protect\citeauthoryear{??}{mic}{2015}]%
        {microsoftdeepcrossdomain}
 \bibinfo{year}{2015}\natexlab{}.
\newblock \showarticletitle{A Multi-View Deep Learning Approach for Cross
  Domain User Modeling in Recommendation Systems}. In \bibinfo{booktitle}{{\em
  Proceedings of the 24th International Conference on World Wide Web}} {\em
  (\bibinfo{series}{WWW '15})}. \bibinfo{publisher}{International World Wide
  Web Conferences Steering Committee}, \bibinfo{address}{Republic and Canton of
  Geneva, Switzerland}, \bibinfo{pages}{278--288}.
\newblock
\showISBNx{978-1-4503-3469-3}
\showDOI{%
\url{https://doi.org/10.1145/2736277.2741667}}


\bibitem[\protect\citeauthoryear{Abdi and Williams}{Abdi and Williams}{2010}]%
        {pca}
\bibfield{author}{\bibinfo{person}{Herv{\'e} Abdi} {and}
  \bibinfo{person}{Lynne~J Williams}.} \bibinfo{year}{2010}\natexlab{}.
\newblock \showarticletitle{Principal component analysis}.
\newblock \bibinfo{journal}{{\em Wiley interdisciplinary reviews: computational
  statistics\/}} \bibinfo{volume}{2}, \bibinfo{number}{4}
  (\bibinfo{year}{2010}), \bibinfo{pages}{433--459}.
\newblock


\bibitem[\protect\citeauthoryear{Cheng, Koc, Harmsen, Shaked, Chandra, Aradhye,
  Anderson, Corrado, Chai, Ispir, Anil, Haque, Hong, Jain, Liu, and Shah}{Cheng
  et~al\mbox{.}}{2016}]%
        {wide-and-deep}
\bibfield{author}{\bibinfo{person}{Heng{-}Tze Cheng}, \bibinfo{person}{Levent
  Koc}, \bibinfo{person}{Jeremiah Harmsen}, \bibinfo{person}{Tal Shaked},
  \bibinfo{person}{Tushar Chandra}, \bibinfo{person}{Hrishi Aradhye},
  \bibinfo{person}{Glen Anderson}, \bibinfo{person}{Greg Corrado},
  \bibinfo{person}{Wei Chai}, \bibinfo{person}{Mustafa Ispir},
  \bibinfo{person}{Rohan Anil}, \bibinfo{person}{Zakaria Haque},
  \bibinfo{person}{Lichan Hong}, \bibinfo{person}{Vihan Jain},
  \bibinfo{person}{Xiaobing Liu}, {and} \bibinfo{person}{Hemal Shah}.}
  \bibinfo{year}{2016}\natexlab{}.
\newblock \showarticletitle{Wide {\&} Deep Learning for Recommender Systems}.
\newblock \bibinfo{journal}{{\em CoRR\/}}  \bibinfo{volume}{abs/1606.07792}
  (\bibinfo{year}{2016}).
\newblock
\showURL{%
\url{http://arxiv.org/abs/1606.07792}}


\bibitem[\protect\citeauthoryear{Covington, Adams, and Sargin}{Covington
  et~al\mbox{.}}{2016}]%
        {youtuberec}
\bibfield{author}{\bibinfo{person}{Paul Covington}, \bibinfo{person}{Jay
  Adams}, {and} \bibinfo{person}{Emre Sargin}.}
  \bibinfo{year}{2016}\natexlab{}.
\newblock \showarticletitle{Deep Neural Networks for YouTube Recommendations}.
  In \bibinfo{booktitle}{{\em Proceedings of the 10th ACM Conference on
  Recommender Systems}}. \bibinfo{address}{New York, NY, USA}.
\newblock


\bibitem[\protect\citeauthoryear{Fayyad and Irani}{Fayyad and Irani}{1993}]%
        {MDL}
\bibfield{author}{\bibinfo{person}{Usama~M. Fayyad} {and}
  \bibinfo{person}{Keki~B. Irani}.} \bibinfo{year}{1993}\natexlab{}.
\newblock \showarticletitle{Multi-Interval Discretization of Continuous-Valued
  Attributes for Classification Learning.}. In \bibinfo{booktitle}{{\em
  IJCAI}}. \bibinfo{pages}{1022--1029}.
\newblock


\bibitem[\protect\citeauthoryear{Herlocker, Konstan, Borchers, and
  Riedl}{Herlocker et~al\mbox{.}}{1999}]%
        {Herlocker1999}
\bibfield{author}{\bibinfo{person}{Jonathan~L. Herlocker},
  \bibinfo{person}{Joseph~A. Konstan}, \bibinfo{person}{Al Borchers}, {and}
  \bibinfo{person}{John Riedl}.} \bibinfo{year}{1999}\natexlab{}.
\newblock \showarticletitle{An Algorithmic Framework for Performing
  Collaborative Filtering}. In \bibinfo{booktitle}{{\em Proceedings of the 22Nd
  Annual International ACM SIGIR Conference on Research and Development in
  Information Retrieval}} {\em (\bibinfo{series}{SIGIR '99})}.
  \bibinfo{publisher}{ACM}, \bibinfo{address}{New York, NY, USA},
  \bibinfo{pages}{230--237}.
\newblock
\showISBNx{1-58113-096-1}
\showDOI{%
\url{https://doi.org/10.1145/312624.312682}}


\bibitem[\protect\citeauthoryear{Hinton and Salakhutdinov}{Hinton and
  Salakhutdinov}{2006}]%
        {deepAutoEncoder}
\bibfield{author}{\bibinfo{person}{G.~E. Hinton} {and} \bibinfo{person}{R.~R.
  Salakhutdinov}.} \bibinfo{year}{2006}\natexlab{}.
\newblock \showarticletitle{Reducing the Dimensionality of Data with Neural
  Networks}.
\newblock \bibinfo{journal}{{\em Science\/}} \bibinfo{volume}{313},
  \bibinfo{number}{5786} (\bibinfo{year}{2006}), \bibinfo{pages}{504--507}.
\newblock


\bibitem[\protect\citeauthoryear{Hu, Koren, and Volinsky}{Hu
  et~al\mbox{.}}{2008}]%
        {implicit}
\bibfield{author}{\bibinfo{person}{Yifan Hu}, \bibinfo{person}{Yehuda Koren},
  {and} \bibinfo{person}{Chris Volinsky}.} \bibinfo{year}{2008}\natexlab{}.
\newblock \showarticletitle{Collaborative Filtering for Implicit Feedback
  Datasets}. In \bibinfo{booktitle}{{\em Proceedings of the 2008 Eighth IEEE
  International Conference on Data Mining}} {\em (\bibinfo{series}{ICDM '08})}.
  \bibinfo{publisher}{IEEE Computer Society}, \bibinfo{address}{Washington, DC,
  USA}, \bibinfo{pages}{263--272}.
\newblock
\showISBNx{978-0-7695-3502-9}
\showDOI{%
\url{https://doi.org/10.1109/ICDM.2008.22}}


\bibitem[\protect\citeauthoryear{Jasper~Snoek and Adams}{Jasper~Snoek and
  Adams}{2012}]%
        {whetlab1}
\bibfield{author}{\bibinfo{person}{Hugo~Larochelle Jasper~Snoek} {and}
  \bibinfo{person}{Ryan~Prescott Adams}.} \bibinfo{year}{2012}\natexlab{}.
\newblock \showarticletitle{Practical Bayesian Optimization of Machine Learning
  Algorithms}.
\newblock \bibinfo{journal}{{\em Advances in Neural Information Processing
  Systems\/}} (\bibinfo{year}{2012}).
\newblock


\bibitem[\protect\citeauthoryear{Karau, Konwinski, Wendell, and Zaharia}{Karau
  et~al\mbox{.}}{2015}]%
        {sparkSVD}
\bibfield{author}{\bibinfo{person}{Holden Karau}, \bibinfo{person}{Andy
  Konwinski}, \bibinfo{person}{Patrick Wendell}, {and} \bibinfo{person}{Matei
  Zaharia}.} \bibinfo{year}{2015}\natexlab{}.
\newblock \bibinfo{booktitle}{{\em Learning Spark: Lightning-Fast Big Data
  Analytics\/} (\bibinfo{edition}{1st} ed.)}.
\newblock \bibinfo{publisher}{O'Reilly Media, Inc.}
\newblock
\showISBNx{1449358624, 9781449358624}


\bibitem[\protect\citeauthoryear{Koren}{Koren}{2008}]%
        {svdpp}
\bibfield{author}{\bibinfo{person}{Yehuda Koren}.}
  \bibinfo{year}{2008}\natexlab{}.
\newblock \showarticletitle{Factorization Meets the Neighborhood: A
  Multifaceted Collaborative Filtering Model}. In \bibinfo{booktitle}{{\em
  Proceedings of the 14th ACM SIGKDD International Conference on Knowledge
  Discovery and Data Mining}} {\em (\bibinfo{series}{KDD '08})}.
  \bibinfo{publisher}{ACM}, \bibinfo{address}{New York, NY, USA},
  \bibinfo{pages}{426--434}.
\newblock
\showISBNx{978-1-60558-193-4}
\showDOI{%
\url{https://doi.org/10.1145/1401890.1401944}}


\bibitem[\protect\citeauthoryear{Krizhevsky, Sutskever, and Hinton}{Krizhevsky
  et~al\mbox{.}}{2012}]%
        {imagenet}
\bibfield{author}{\bibinfo{person}{Alex Krizhevsky}, \bibinfo{person}{Ilya
  Sutskever}, {and} \bibinfo{person}{Geoffrey~E Hinton}.}
  \bibinfo{year}{2012}\natexlab{}.
\newblock \showarticletitle{Imagenet classification with deep convolutional
  neural networks}. In \bibinfo{booktitle}{{\em Advances in neural information
  processing systems}}. \bibinfo{pages}{1097--1105}.
\newblock


\bibitem[\protect\citeauthoryear{Mandelbaum and Shalev}{Mandelbaum and
  Shalev}{2016}]%
        {w2vdownstream}
\bibfield{author}{\bibinfo{person}{Amit Mandelbaum} {and} \bibinfo{person}{Adi
  Shalev}.} \bibinfo{year}{2016}\natexlab{}.
\newblock \showarticletitle{Word Embeddings and Their Use In Sentence
  Classification Tasks}.
\newblock \bibinfo{journal}{{\em CoRR\/}}  \bibinfo{volume}{abs/1610.08229}
  (\bibinfo{year}{2016}).
\newblock
\showeprint[arxiv]{1610.08229}
\showURL{%
\url{http://arxiv.org/abs/1610.08229}}


\bibitem[\protect\citeauthoryear{Martinsson, Rokhlin, and Tygert}{Martinsson
  et~al\mbox{.}}{2011}]%
        {randomSVD}
\bibfield{author}{\bibinfo{person}{Per-Gunnar Martinsson},
  \bibinfo{person}{Vladimir Rokhlin}, {and} \bibinfo{person}{Mark Tygert}.}
  \bibinfo{year}{2011}\natexlab{}.
\newblock \showarticletitle{A randomized algorithm for the decomposition of
  matrices}.
\newblock \bibinfo{journal}{{\em Applied and Computational Harmonic
  Analysis\/}} \bibinfo{volume}{30}, \bibinfo{number}{1}
  (\bibinfo{year}{2011}), \bibinfo{pages}{47 -- 68}.
\newblock
\showISSN{1063-5203}
\showDOI{%
\url{https://doi.org/10.1016/j.acha.2010.02.003}}


\bibitem[\protect\citeauthoryear{Mikolov, Chen, Corrado, and Dean}{Mikolov
  et~al\mbox{.}}{2013}]%
        {word2vec}
\bibfield{author}{\bibinfo{person}{Tomas Mikolov}, \bibinfo{person}{Kai Chen},
  \bibinfo{person}{Greg Corrado}, {and} \bibinfo{person}{Jeffrey Dean}.}
  \bibinfo{year}{2013}\natexlab{}.
\newblock \showarticletitle{Efficient Estimation of Word Representations in
  Vector Space}.
\newblock \bibinfo{journal}{{\em CoRR\/}}  \bibinfo{volume}{abs/1301.3781}
  (\bibinfo{year}{2013}).
\newblock
\showeprint[arxiv]{1301.3781}
\showURL{%
\url{http://arxiv.org/abs/1301.3781}}


\bibitem[\protect\citeauthoryear{Nair and Hinton}{Nair and Hinton}{2010}]%
        {relu}
\bibfield{author}{\bibinfo{person}{Vinod Nair} {and}
  \bibinfo{person}{Geoffrey~E. Hinton}.} \bibinfo{year}{2010}\natexlab{}.
\newblock \showarticletitle{Rectified Linear Units Improve Restricted Boltzmann
  Machines}. In \bibinfo{booktitle}{{\em Proceedings of the 27th International
  Conference on International Conference on Machine Learning}} {\em
  (\bibinfo{series}{ICML'10})}.
\newblock


\bibitem[\protect\citeauthoryear{Perozzi, Al{-}Rfou, and Skiena}{Perozzi
  et~al\mbox{.}}{2014}]%
        {deepwalk}
\bibfield{author}{\bibinfo{person}{Bryan Perozzi}, \bibinfo{person}{Rami
  Al{-}Rfou}, {and} \bibinfo{person}{Steven Skiena}.}
  \bibinfo{year}{2014}\natexlab{}.
\newblock \showarticletitle{DeepWalk: Online Learning of Social
  Representations}.
\newblock \bibinfo{journal}{{\em CoRR\/}}  \bibinfo{volume}{abs/1403.6652}
  (\bibinfo{year}{2014}).
\newblock
\showeprint[arxiv]{1403.6652}
\showURL{%
\url{http://arxiv.org/abs/1403.6652}}


\bibitem[\protect\citeauthoryear{Quionero-Candela, Sugiyama, Schwaighofer, and
  Lawrence}{Quionero-Candela et~al\mbox{.}}{2009}]%
        {datasetshift}
\bibfield{author}{\bibinfo{person}{Joaquin Quionero-Candela},
  \bibinfo{person}{Masashi Sugiyama}, \bibinfo{person}{Anton Schwaighofer},
  {and} \bibinfo{person}{Neil~D. Lawrence}.} \bibinfo{year}{2009}\natexlab{}.
\newblock \bibinfo{booktitle}{{\em Dataset Shift in Machine Learning}}.
\newblock \bibinfo{publisher}{The MIT Press}.
\newblock
\showISBNx{0262170051, 9780262170055}


\bibitem[\protect\citeauthoryear{Salakhutdinov and Mnih}{Salakhutdinov and
  Mnih}{2007}]%
        {pmf}
\bibfield{author}{\bibinfo{person}{Ruslan Salakhutdinov} {and}
  \bibinfo{person}{Andriy Mnih}.} \bibinfo{year}{2007}\natexlab{}.
\newblock \showarticletitle{Probabilistic Matrix Factorization}. In
  \bibinfo{booktitle}{{\em Proceedings of the 20th International Conference on
  Neural Information Processing Systems}} {\em (\bibinfo{series}{NIPS'07})}.
  \bibinfo{publisher}{Curran Associates Inc.}, \bibinfo{address}{USA},
  \bibinfo{pages}{1257--1264}.
\newblock
\showISBNx{978-1-60560-352-0}
\showURL{%
\url{http://dl.acm.org/citation.cfm?id=2981562.2981720}}


\bibitem[\protect\citeauthoryear{Sarwar, Karypis, Konstan, and Riedl}{Sarwar
  et~al\mbox{.}}{2002}]%
        {foldingin}
\bibfield{author}{\bibinfo{person}{Badrul Sarwar}, \bibinfo{person}{George
  Karypis}, \bibinfo{person}{Joseph Konstan}, {and} \bibinfo{person}{John
  Riedl}.} \bibinfo{year}{2002}\natexlab{}.
\newblock \showarticletitle{Incremental singular value decomposition algorithms
  for highly scalable recommender systems}. In \bibinfo{booktitle}{{\em Fifth
  International Conference on Computer and Information Science}}. Citeseer,
  \bibinfo{pages}{27--28}.
\newblock


\bibitem[\protect\citeauthoryear{Schelter, Boden, Schenck, Alexandrov, and
  Markl}{Schelter et~al\mbox{.}}{2013}]%
        {mapreduceALS}
\bibfield{author}{\bibinfo{person}{Sebastian Schelter},
  \bibinfo{person}{Christoph Boden}, \bibinfo{person}{Martin Schenck},
  \bibinfo{person}{Alexander Alexandrov}, {and} \bibinfo{person}{Volker
  Markl}.} \bibinfo{year}{2013}\natexlab{}.
\newblock \showarticletitle{Distributed Matrix Factorization with Mapreduce
  Using a Series of Broadcast-joins}. In \bibinfo{booktitle}{{\em Proceedings
  of the 7th ACM Conference on Recommender Systems}} {\em
  (\bibinfo{series}{RecSys '13})}. \bibinfo{publisher}{ACM},
  \bibinfo{address}{New York, NY, USA}, \bibinfo{pages}{281--284}.
\newblock
\showISBNx{978-1-4503-2409-0}
\showDOI{%
\url{https://doi.org/10.1145/2507157.2507195}}


\bibitem[\protect\citeauthoryear{Sculley, Holt, Golovin, Davydov, Phillips,
  Ebner, Chaudhary, and Young}{Sculley et~al\mbox{.}}{2014}]%
        {creditcard}
\bibfield{author}{\bibinfo{person}{D. Sculley}, \bibinfo{person}{Gary Holt},
  \bibinfo{person}{Daniel Golovin}, \bibinfo{person}{Eugene Davydov},
  \bibinfo{person}{Todd Phillips}, \bibinfo{person}{Dietmar Ebner},
  \bibinfo{person}{Vinay Chaudhary}, {and} \bibinfo{person}{Michael Young}.}
  \bibinfo{year}{2014}\natexlab{}.
\newblock \showarticletitle{Machine Learning: The High Interest Credit Card of
  Technical Debt}. In \bibinfo{booktitle}{{\em SE4ML: Software Engineering for
  Machine Learning (NIPS 2014 Workshop)}}.
\newblock


\bibitem[\protect\citeauthoryear{Vincent, Larochelle, Bengio, and
  Manzagol}{Vincent et~al\mbox{.}}{2008}]%
        {autoencoder}
\bibfield{author}{\bibinfo{person}{Pascal Vincent}, \bibinfo{person}{Hugo
  Larochelle}, \bibinfo{person}{Yoshua Bengio}, {and}
  \bibinfo{person}{Pierre-Antoine Manzagol}.} \bibinfo{year}{2008}\natexlab{}.
\newblock \showarticletitle{Extracting and Composing Robust Features with
  Denoising Autoencoders}. In \bibinfo{booktitle}{{\em Proceedings of the 25th
  International Conference on Machine Learning}} {\em (\bibinfo{series}{ICML
  '08})}. \bibinfo{publisher}{ACM}, \bibinfo{address}{New York, NY, USA},
  \bibinfo{pages}{1096--1103}.
\newblock
\showISBNx{978-1-60558-205-4}
\showDOI{%
\url{https://doi.org/10.1145/1390156.1390294}}


\bibitem[\protect\citeauthoryear{Weinberger, Dasgupta, Langford, Smola, and
  Attenberg}{Weinberger et~al\mbox{.}}{2009}]%
        {featurehashing}
\bibfield{author}{\bibinfo{person}{Kilian Weinberger}, \bibinfo{person}{Anirban
  Dasgupta}, \bibinfo{person}{John Langford}, \bibinfo{person}{Alex Smola},
  {and} \bibinfo{person}{Josh Attenberg}.} \bibinfo{year}{2009}\natexlab{}.
\newblock \showarticletitle{Feature Hashing for Large Scale Multitask
  Learning}. In \bibinfo{booktitle}{{\em Proceedings of the 26th Annual
  International Conference on Machine Learning}} {\em (\bibinfo{series}{ICML
  '09})}. \bibinfo{publisher}{ACM}, \bibinfo{address}{New York, NY, USA},
  \bibinfo{pages}{1113--1120}.
\newblock
\showISBNx{978-1-60558-516-1}
\showDOI{%
\url{https://doi.org/10.1145/1553374.1553516}}


\bibitem[\protect\citeauthoryear{Wu, Fisch, Chopra, Adams, Bordes, and
  Weston}{Wu et~al\mbox{.}}{2017}]%
        {starspace}
\bibfield{author}{\bibinfo{person}{Ledell Wu}, \bibinfo{person}{Adam Fisch},
  \bibinfo{person}{Sumit Chopra}, \bibinfo{person}{Keith Adams},
  \bibinfo{person}{Antoine Bordes}, {and} \bibinfo{person}{Jason Weston}.}
  \bibinfo{year}{2017}\natexlab{}.
\newblock \showarticletitle{StarSpace: Embed All The Things!}
\newblock \bibinfo{journal}{{\em CoRR\/}}  \bibinfo{volume}{abs/1709.03856}
  (\bibinfo{year}{2017}).
\newblock
\showeprint[arxiv]{1709.03856}
\showURL{%
\url{http://arxiv.org/abs/1709.03856}}


\bibitem[\protect\citeauthoryear{Xiong}{Xiong}{2007}]%
        {sparseclassification}
\bibfield{author}{\bibinfo{person}{Tao Xiong}.}
  \bibinfo{year}{2007}\natexlab{}.
\newblock {\em \bibinfo{title}{Classification Methods for High-dimensional
  Sparse Data}}.
\newblock \bibinfo{thesistype}{Ph.D. Dissertation}.
  \bibinfo{address}{Minneapolis, MN, USA}.
\newblock Advisor(s) Cherkassky, Vladimir S.
\newblock
\newblock
\shownote{AAI3250170.}


\bibitem[\protect\citeauthoryear{Yadlowsky, Nakkarin, Wang, Sharma, and
  Ghaoui}{Yadlowsky et~al\mbox{.}}{2014}]%
        {sparsetextcorpora}
\bibfield{author}{\bibinfo{person}{Steve Yadlowsky}, \bibinfo{person}{Preetum
  Nakkarin}, \bibinfo{person}{Jingyan Wang}, \bibinfo{person}{Rishi Sharma},
  {and} \bibinfo{person}{Laurent~El Ghaoui}.} \bibinfo{year}{2014}\natexlab{}.
\newblock \showarticletitle{Iterative Hard Thresholding for Keyword Extraction
  from Large Text Corpora}. In \bibinfo{booktitle}{{\em 13th International
  Conference on Machine Learning and Applications, {ICMLA} 2014, Detroit, MI,
  USA, December 3-6, 2014}}. \bibinfo{pages}{588--593}.
\newblock
\showDOI{%
\url{https://doi.org/10.1109/ICMLA.2014.101}}


\end{thebibliography}
